\documentclass{article}
\usepackage[dvips]{graphicx}
\usepackage{latexsym}
\usepackage[fleqn]{amsmath}

\newtheorem{theorem}{Theorem}

\newtheorem{proposition}{Proposition}

\def\squarebox#1{\hbox to #1{\hfill\vbox to #1{\vfill}}}
\def\qed{\hspace*{\fill}
        \vbox{\hrule\hbox{\vrule\squarebox{.667em}\vrule}\hrule}\smallskip}
\newenvironment{proof}{\begin{trivlist}
  \item[\hspace{\labelsep}{\em\noindent Proof.~}]
  }{\qed\end{trivlist}}

\begin{document}

\title{Quantum Identification of Boolean Oracles}

\author{
Andris Ambainis$^{*}$$\qquad$
Kazuo Iwama$^{\dag, \ddag}$$\qquad$
Akinori Kawachi$^{\dag, \ddag}$\\
Hiroyuki Masuda$^{\dag, \ddag}$$\quad$
Raymond H. Putra$^{\dag, \ddag}$$\quad$
Shigeru Yamashita$^{\S}$\\
\\
$^{*}$School of Mathematics, Institute for Advanced Study\\
 {\tt ambainis@ias.edu}\\
$^{\dag}$Quantum Computation and Information,  ERATO,\\
Japan Science and Technology Corporation (JST)\\
$^{\ddag}$Graduate School of Informatics,  Kyoto University \\
{\tt \{iwama, kawachi, hiroyuki, raymond\}@kuis.kyoto-u.ac.jp}\\
$^{\S}$Graduate School of Information Science, Nara Institute of Science and Technology\\
{\tt ger@is.aist-nara.ac.jp}
}
\date{}

\maketitle

\begin{abstract}
The oracle identification problem (OIP) is, 
given a set $S$ of $M$ Boolean oracles out of $2^{N}$ ones, 
to determine which oracle in $S$ is the current black-box oracle. 
We can exploit the information that candidates of the current oracle is 
restricted to $S$. 
The OIP contains several concrete problems such as the original 
Grover search and the Bernstein-Vazirani problem. 
Our interest is in the quantum query complexity, 
for which we present several upper and lower bounds. 
They are quite general and mostly optimal: 
(i) The query complexity of OIP is $O(\sqrt{N\log M \log N}\log\log M)$ 
for {\it any} $S$ such that $M = |S| > N$, 
which is better than the obvious bound $N$ if $M < 2^{N/\log^{3}N}$. 
(ii) It is $O(\sqrt{N})$ for {\it any} $S$ if $|S| = N$, which includes 
the upper bound for the Grover search as a special case. 
(iii) For a wide range of oracles ($|S| = N$) 
such as random oracles and balanced oracles, the query complexity is $\Theta(\sqrt{N/K})$, 
where $K$ is a simple parameter determined by $S$.
\end{abstract}

\section{Introduction}
An {\it oracle} is given as a Boolean function of $n$ variables, 
denoted by $f(x_{0},\ldots,x_{n-1})$, 
and so there are $2^{2^{n}}$ (or $2^{N}$ for $N = 2^{n}$) different oracles. 
An {\it oracle computation} is, 
given a specific oracle $f$ which we do not know, to determine, 
through queries to the oracle, 
whether or not $f$ satisfies a certain property. 
Note that $f$ has $N$ black-box $0/1$-values, 
$f(0,\ldots,0)$ through $f(1,\ldots,1)$. 
($f(0,\ldots,0)$ is also denoted as $f(0)$, 
$f(1,\ldots,1)$ as $f(N-1)$, 
and similarly for an intermediate $f(j)$.) 
So, in other words, 
we are asked whether or not these $N$ bits satisfy the property. 
There are many interesting such properties: 
For example, it is called {\it OR} if the question is whether all the $N$ bits are $0$ 
and {\it Parity} if the question is whether the $N$ bits include an even number of $1$'s. 
The most general question (or job in this case) is to obtain all the $N$ bits. 
Our complexity measure is the so-called {\it query complexity}, i.e., 
the number of oracle calls, to get a right answer with bounded error. 
Note that the trivial upper bound is $N$ since we can tell all the $N$ bits by asking $f(0)$ through $f(N-1)$. 
If we use a classical computer, this $N$ is also a lower bound in most cases. 
If we use a quantum computer, however, several interesting speedups are obtained. 
For example, the previous three problems have a (quantum) query complexity of $O(\sqrt{N})$, 
$\frac{N}{2}$ and $\frac{N}{2}+\sqrt{N}$, respectively \cite{Gro96,BBHT96,FGGS98,Dam98}. 

In this paper, we discuss the following problem which we call the {\it oracle identification} problem: 
We are given a set $S$ of $M$ different oracles out of the $2^{N}$ ones for which we have the complete information 
(i.e., for each of the $2^{N}$ oracles, we know whether it is in $S$ or not). 
Now we are asked to determine which oracle in $S$ is currently in the black-box. 
A typical example is the Grover search \cite{Gro96} where $S = \{f_{0},\ldots,f_{N-1}\}$ 
and $f_{i}(j) = 1$ iff $i = j$. 
(Namely, exactly one bit among the $N$ bits is $1$ in each oracle in $S$. 
Finding its position is equivalent to identifying the oracle itself.) 
It is well-known that its query complexity is $\Theta(\sqrt{N})$. 
Another example is the so-called Bernstein-Vazirani problem \cite{Bernstein97} where $S = \{f_{0},\ldots,f_{N-1}\}$ 
and $f_{i}(j) = 1$ iff the inner product of $i$ and $j$ (mod $2$) is 1. 
A little surprisingly, its query complexity is just one. 

Thus the oracle identification problem is a promise version of the oracle computation problem. 
For both oracle computation and oracle identification problems, 
\cite{Amb02} developed a very general method for proving their lower bounds of the query complexity. 
Also, many nontrivial upper bounds are known as mentioned above. 
However all those upper bounds are for specific problems such as the Grover search; 
no general upper bounds for a wide class of problems have been known so far. 

{\bf Our Contribution.}
In this paper, we give  general upper and lower bounds for the oracle identification problem. 
More concretely we prove: 
(i)~The query complexity of the oracle identification for {\it any} oracle set $S$ is
$O(\sqrt{N\log M \log N}\log\log M)$ if $|S| = M > N$. 
(ii)~It is $O(\sqrt{N})$ for {\it any} $S$ if $|S| = N$. 
(iii)~For a wide range of oracles ($M = N$) such as random oracles and balanced oracles, 
the query complexity is $\Theta(\sqrt{\frac{N}{K}})$, 
where $K$ is a parameter determined by $S$. 
The bound in (i) is better than the obvious bound $N$ if $M < 2^{N/\log^{3}N}$.
Both algorithms for (i) and (ii) are quite tricky, 
and the result (ii) includes the upper bound for the Grover search as a special case. 
Result (i) is almost optimal, and results (ii) and (iii) are optimal; 
to prove their optimality we introduce a general lower bound theorem whose statement is 
simpler than that of \cite{Amb02}. 

{\bf Related Results.} Query complexity has constantly been one of the
central topics in quantum computation; to cover everything is obviously impossible. 
For the upper bounds of the query complexity, the most significant result is 
due to \cite{Gro96}, known as the Grover search, 
which also derived many applications and extensions \cite{BBBGL98,BBHT96,CK97,Gro98,Gro00}. 
In particular, some results showed efficient quantum algorithms by combining the Grover search 
with other (quantum and classical) techniques.
For example, quantum counting algorithm \cite{BHMT00} gives an approximate counting method by combining the Grover search with 
the quantum Fourier transformation, and quantum algorithms for the claw-finding and the element distinctness problems
\cite{BDHHMSW01} also exploit classical random sampling and sorting. 
Most recently, \cite{Amb03b} developed an optimal quantum algorithm with $O(N^{2/3})$ queries for element distinctness 
problem, which makes use of quantum walk and matches to the lower bounds shown by \cite{Shi02}.
\cite{MSS03} used the element distinctness algorithm to design a quantum algorithm for finding triangles in a graph.
\cite{AA03} also showed an efficient quantum search algorithm for spacial regions
based on recursive Grover search, which is applicable to some geometrically structured problems such as search on a 2-D grid.

On the lower-bound side, there are two popular techniques to derive 
quantum lower bounds, i.e., the polynomial method and the quantum adversary method. 
The polynomials method was firstly introduced for 
quantum computation by \cite{BBCMW98} who borrowed the idea from the
classical counterpart. For example, it was shown that for
bounded error cases, evaluations of $AND$ and $OR$ functions need
$\Theta(\sqrt{N})$ number of queries, while parity and majority
functions at least $N/2$ and $\Theta(N)$, respectively. 
Recently, \cite{Aar02, Shi02} used the polynomials method to show the lower bounds
for the collisions and element distinctness problems.

The classical adversary method was used in \cite{BBBV97,Vaz98}, which is also called the hybrid argument. 
Their method can be used, for example, to show the lower bound of the Grover search. 
As mentioned above, \cite{Amb02} introduced a quite general method, 
which is known as the quantum adversary argument, for obtaining lower bounds of various problems, e.g., the Grover
search, AND of ORs and inverting a permutation. 
\cite{BS02} recently established a lower bound of $\Omega(\sqrt{N})$ 
on the bounded-error quantum query complexity of read-once Boolean functions by extending \cite{Amb02}.
\cite{BSS03} generalized the quantum adversary method respectively from the aspect of semidefinite programming, 
and \cite{LM04} generalized the method from Kolmogorov complexity perspective.
Furthermore, \cite{CML03,Aar03} showed the lower bounds for graph connectivity and local search problem respectively 
using the quantum adversary method.
\cite{Amb03a} also gave a comparison between the quantum adversary method and the polynomial method.

\section{Formalization}

Our model is basically the same as standard ones (see e.g., \cite{Amb02}). 
For a Boolean function $f(x_0, \ldots, x_{n-1})$ of $n$ variables, an {\it oracle} maps
$|x_{0}, \ldots, x_{n-1}\rangle|b\rangle$ to 
$(-1)^{b \cdot f(x_{0}, \ldots, x_{n-1})}|x_{0}, \ldots, x_{n-1}\rangle|b\rangle$. 
A {\it quantum computation} is a sequence of unitary transformations 
$U_{0} \to O \to U_{1} \to O \to \cdots \to O \to U_{t}$, where $O$ is a single 
oracle call against our {\it black-box oracle} 
(sometimes called an {\it input oracle}), 
and $U_{j}$ may be any unitary transformation without oracle calls. 
The above computation sequence involves $t$ oracle calls, 
which is our measure of the complexity 
({\it the query complexity}). 
Let $N = 2^{n}$ and hence there are $2^{N}$ different oracles. 

Our problem is called the {\it Oracle Identification Problem} ({\it OIP}). 
An OIP is given as an infinite sequence 
$S_{1}, S_{2}, S_{4}, \ldots, S_{N}, \ldots$. 
Each $S_{N}$ ($N = 2^{n}, n = 0, 1, \ldots$) is a set of oracles 
(Boolean functions with $n$ variables) 
whose size, $|S_{N}|$, is denoted by $M$ ($\le 2^{N}$). 
A (quantum) algorithm $A$ which solves the OIP is a quantum computation as given above. 
$A$ has to determine which oracle ($\in S_{N}$) is the current input oracle with bounded error. 
If $A$ needs at most $g(N)$ oracle calls, we say that the query complexity of $A$ is $g(N)$. 
It should be noted that $A$ knows the set $S_{N}$ completely; 
what is unknown for $A$ is the current input oracle.

For example, the Grover search is an OIP whose $S_{N}$ contains $N$ (i.e., $M = N$) Boolean functions 
$f_{1}, \ldots, f_{N}$ such that
$$
f_{i}(j) = 1 \quad \mbox{iff} \quad i=j.
$$
Note that $f(j)$ means $f(a_{0}, a_{1}, \ldots, a_{n-1})$ ($a_{i} =$ $0$ or $1$) 
such that $a_{0}, \ldots, a_{n-1}$ is the binary representation of the number $j$. 
Note that $S_{N}$ is given as a $N \times M$ Boolean matrix. 
More formally, the entry at row $i$ ($0 \le i \le M-1$) and column $j$ ($0 \le j \le N-1$) shows $f_{i}(j)$. 
Fig.~1 shows such a matrix of the Grover search for $N = M = 16$. 
Each row corresponds to each oracle in $S_{N}$ and each column to its Boolean value.
Fig.~2 shows another famous example given by an $N \times N$ matrix, which is called the 
Bernstein-Vazirani problem \cite{Bernstein97}. 
It is well known that there is an algorithm whose query complexity is just one for this problem \cite{Bernstein97}.

\begin{figure}[h]
\begin{minipage}{.50\linewidth}
\begin{center}
\resizebox{7.5cm}{!}{ 
\includegraphics*{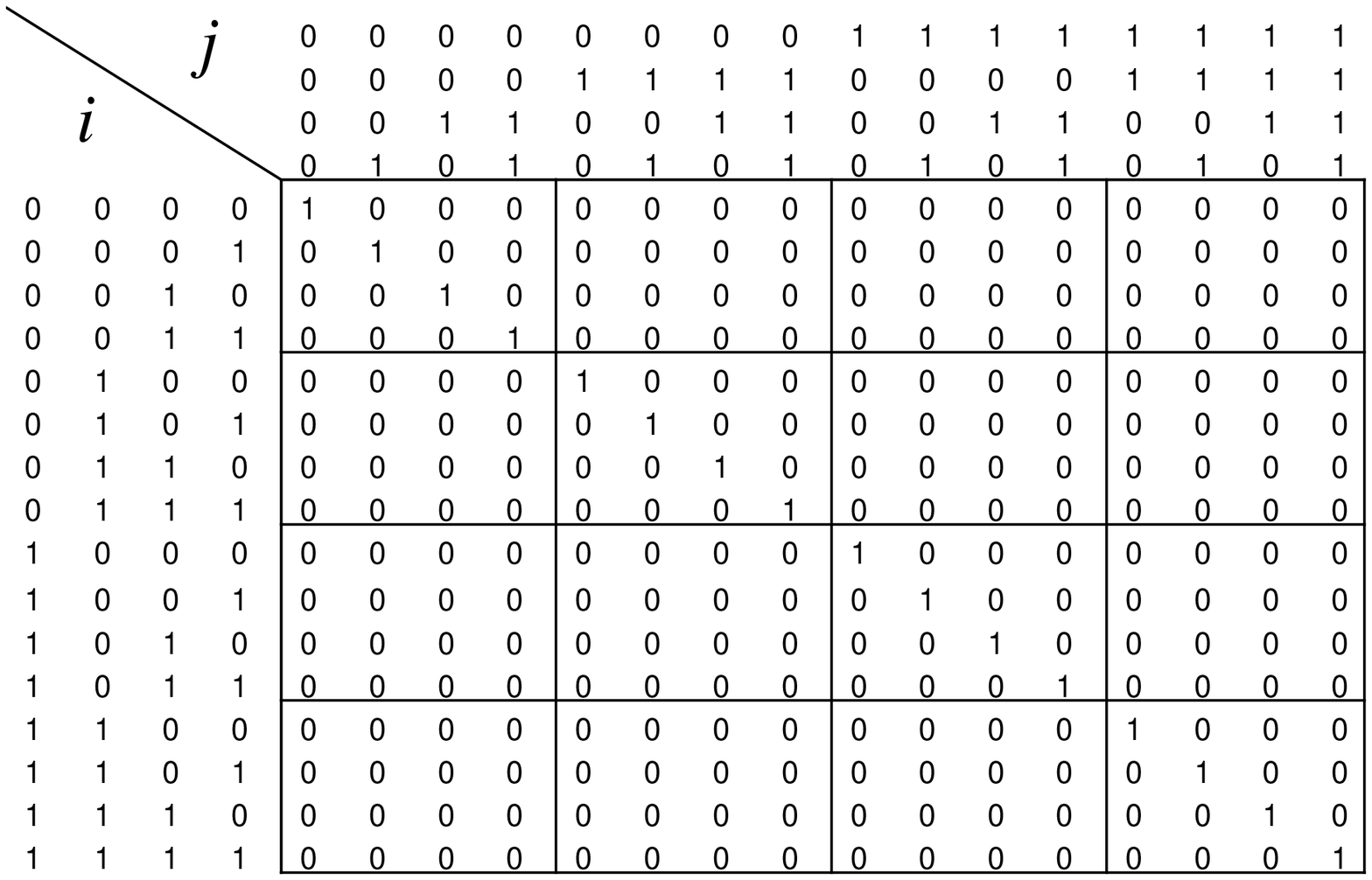} 
}
\end{center}
\caption{$f_{i}(j) = 1 \quad \mbox{iff} \quad i=j$}
\end{minipage}
\begin{minipage}{.50\linewidth}
\begin{center}
\resizebox{7.5cm}{!}{ 
\includegraphics*{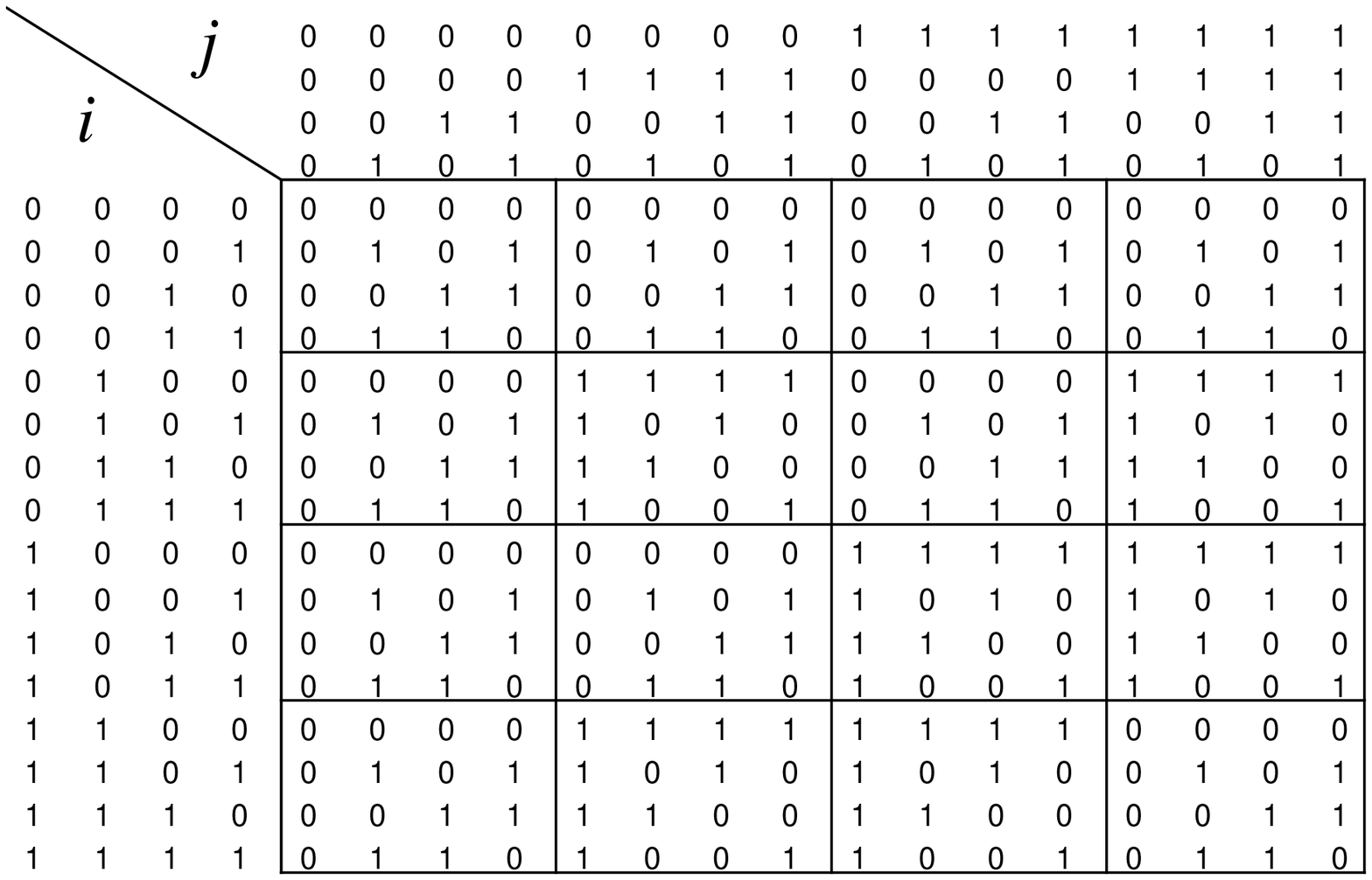} 
}
\end{center}
\caption{$f_{i}(j) = i\cdot j = \sum_x i_x \cdot j_x\,\,\mod 2$}
\end{minipage}
\end{figure}

As described in the previous section, there are several similar (but different subtly) settings. 
For example, the problem in \cite{NW99,Amb02} is given as a matrix 
which includes all the rows (oracles) each of which contains $N/2$ $1$'s or $(1/2+\varepsilon)N$ $1$'s for $\varepsilon > 0$. 
We do not have to identify the current input oracle itself but have only to answer whether the current oracle has $N/2$ $1$'s or not. 
(The famous Deutsch-Jozsa problem \cite{DJ92} is its special case.)
The $l$-target Grover search is given as a matrix consisting of all (or a part of) the rows containing $l$ $1$'s. 
Again we do not have to identify the current input oracle 
but have to answer with a column which has value $1$ in the current input. 
Fig.~3 shows an example, where each row contains $N/2 + 1$ ones. 
One can see that the multi-target Grover search is easy 
($O(1)$ queries are enough since we have roughly one half $1$'s), 
but identifying the input oracle itself is much harder. 

\begin{figure}[h]
\begin{center}
\resizebox{6cm}{!}{ 
\includegraphics*{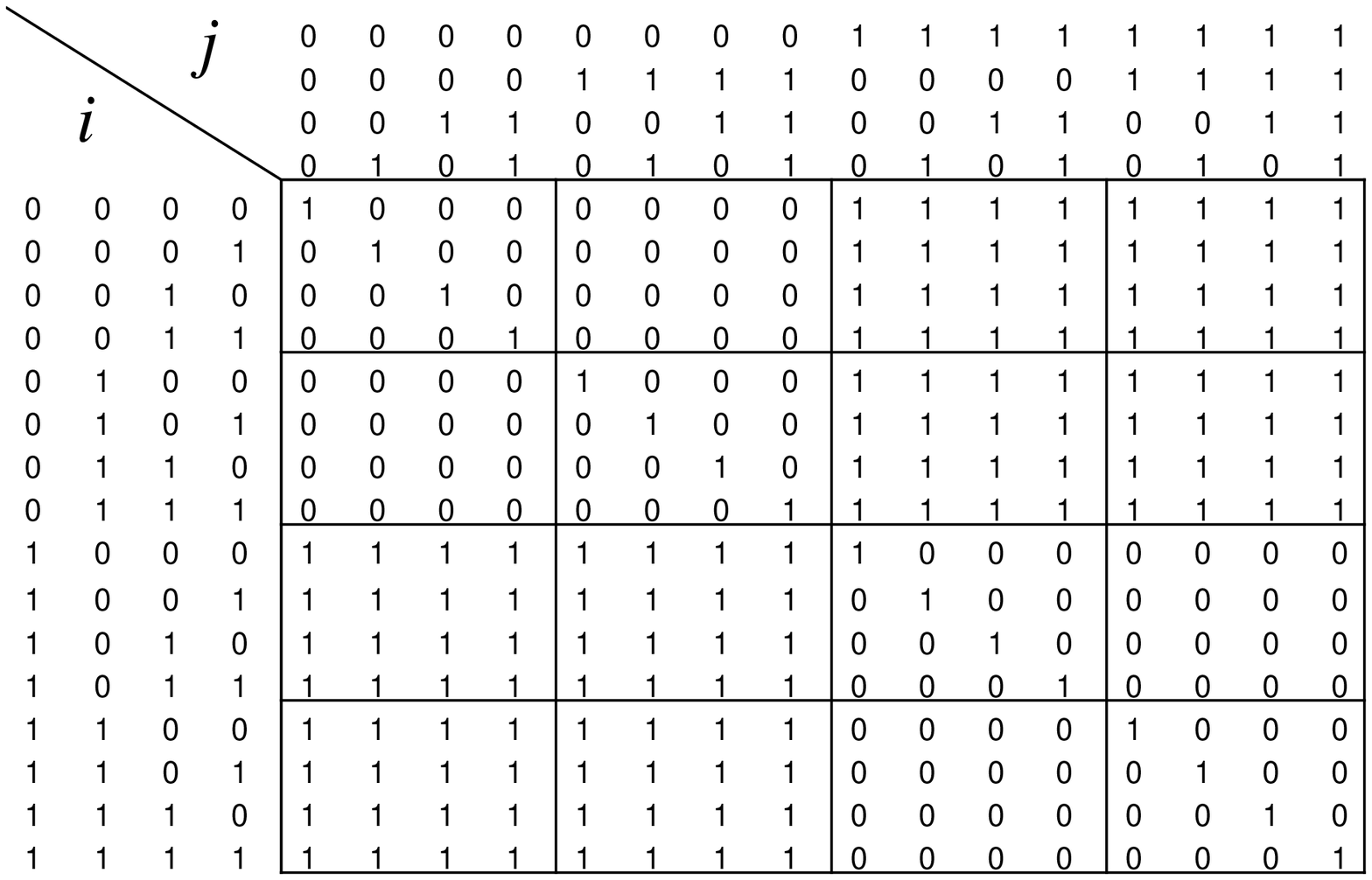} 
}
\caption{Harder case}
\end{center}
\end{figure}

\cite{Amb02} gave a very general lower bounds for oracle computation. 
When applying to the OIP (the original statement is more general), it claims the following: 
\begin{proposition}\label{ambainis}
Let $S_N$ be a given set of oracles,
and $X, Y$ be two disjoint subsets of $S_N$.
Let $R \subset X \times Y$ be such that \\
\hspace{6mm}1. For every $f_{a} \in X$, there exist at least $m$ 
different $f_{b} \in Y$ such that $(f_{a},f_{b}) \in R$. \\
\hspace{6mm}2. For every $f_{b}\in Y$, there exist at least $m'$ 
different $f_{a} \in X$ such that $(f_{a},f_{b}) \in R$. \\
Let $l_{f_{a},i}$ be the number of $f_{b} \in Y$ such that $(f_{a},f_{b}) \in R$ 
and $f_{a}(i) \neq f_{b}(i)$ and 
$l_{f_{b},i}$ be the number of $f_{a} \in X$ such that $(f_{a},f_{b}) \in R$ and 
$f_{a}(i) \neq f_{b}(i)$. 
Let $l_{max}$ be the maximum of $l_{f_{a},i}l_{f_{b},i}$ over all 
$(f_{a},f_{b}) \in R$ and $i \in \{0, \ldots, N-1\}$ such that $f_{a}(i) \neq f_{b}(i)$. 
Then, the query complexity for $S_N$ is 
$\Omega\left(\sqrt{\frac{mm'}{l_{max}}}\right)$. 
\end{proposition}

In this paper, we always assume that $M \ge N$. If $M \le N/2$, then we can select $M$ columns out of the $N$ ones 
while keeping the uniqueness property of each oracle. 
Then by changing the state space from $n$ bits to at most $n-1$ bits, 
we have a new $M \times M$ matrix, i.e., a smaller OIP problem. 

\section{General Upper Bounds}
As mentioned in the previous section, we have a general lower bound for the OIP. 
But we do not know any nontrivial general upper bounds. 
In this section, we give two general upper bounds for the case that $M >N$ and for the case that $M=N$. 
The former is almost tight as described after the theorem, and the latter includes the upper bound for 
the Grover search as a special case. 
An $N \times M$ OIP denotes an OIP 
whose $S_{N}$ (or simply $S$ by omitting the subscript) is 
given as an $N \times M$ matrix as described in the previous section. 
Before proving the theorems, we introduce a convenient technique called a {\it Column Flip}. 

{\bf Column Flip.}
Suppose that $S$ is any $N \times M$ matrix (a set of $M$ oracles). 
Then any quantum computation for $S$ can be transformed into a quantum computation for an $N \times M$ 
matrix $S^{'}$ such that the number of $1$'s is less than or equal to the number of $0$'s in every column. 
(We say that such a matrix is {\it $1$-sensitive}.) 
The reason is straightforward. 
If some column in $S$ holds more $1$'s than $0$'s, then we ``flip" all the values. 
Of course we have to change the current oracle into the new ones 
but this can be easily done by adding an extra circuit to the output of the oracle. 

\begin{theorem}\label{N_times_M_U}
The query complexity of any $N \times M$ OIP is 
$O(\sqrt{N\log M \log N}\log\log M)$ if $M > N$.
\end{theorem}
\begin{proof}
To see the idea, we first prove an easier bound, i.e., $O(\sqrt{N}\log{M}\log\log{M})$. 
(Since $M$ can be an exponential function in $N$, this bound is significantly worse than that of the theorem.) 
If necessary, we convert the given matrix $S$ to be $1$-sensitive by Column Flip. 
Then, just apply the Grover search against the input oracle. 
If we get a column $j$ (the input oracle has $1$ there), then we can eliminate all the rows having $0$ in that column. 
The number of such removed rows is at least one half by the $1$-sensitivity. 
Just repeat this 
(including the conversion to $1$-sensitive matrices)
until the number of rows becomes $1$, 
which needs $O(\log{M})$ rounds. Each Grover Search needs $O(\sqrt{N})$ oracle calls. 
Since we perform many Grover searches, the $\log\log{M}$ term is added to take care of the success probability. 

In this algorithm we counted $O(\sqrt{N})$ oracle calls for the Grover search, which is the target of our improvement. 
More precisely, our algorithm is the following quantum procedure.
Let $S = \{ f_0,...,f_{M-1} \}$ be the given $N\times M$ matrix:

{\it Step 1.}~Let $Z \subseteq S$ be a set of candidate oracles 
(or equivalently an $N \times M$ matrix each row of 
which corresponds to each oracle). Set $Z = S$ initially.

{\it Step 2.}~Repeat Steps 3-6 until $|Z| = 1$.

{\it Step 3.}~Convert $Z$ into $1$-sensitive matrix. 

{\it Step 4.}~Compute the largest integer $K$ such that at least one half rows of $Z$ contain $K$ 1's or more.
(This can be done simply by sorting the rows of $Z$ with the number of 1's.)

{\it Step 5.}~For the current (modified) oracle, perform the multi-target Grover search \cite{BBHT96}
where we set $\frac{9}{2}\sqrt{N/K}$ to the maximum number of oracle calls.
Iterate this Grover search $\log\log M$ times (to increase the success probability).

{\it Step 6.}~If we succeeded in finding 1 by the Grover search in the previous step, i.e., a column $j$ such that 
the current oracle actually has 1 in that column, then eliminate all the rows of $Z$ having 0 in their column $j$.
(Let $Z$ be this reduced matrix.) Otherwise eliminate all the rows of $Z$ having at least $K$ 1's.

Now we estimate the number of oracle calls in this algorithm.
Let $M_r$ and $K_r$ be the number of the rows of $Z$ and the value of $K$ in the $r$-th repetition respectively. Initially, $M_1 = M$.
Note that the number of the rows of $Z$ becomes $|Z|/2$ or less after Step 6, i.e., $M_{r+1}\leq M_r/2$ 
even if the Grover search is successful or not in Step 5 
since the number of 1's in each column of the modified matrix is less than $|Z|/2$ and the number of 
the rows which have at least $K$ 1's is $|Z|/2$ or more. 
Assuming that we need the $T$ repetitions to identify the current input oracle, the total number of the oracle calls is
$$
\frac{9}{2}\left(\sqrt{\frac{N}{K_1}} + \cdots + \sqrt{\frac{N}{K_T}} \right)\log\log M.
$$
We estimate the lower bounds of $K_r$. Note that there are no identical rows in $Z$ and the number of possible rows 
that contain at most $K_r$ 1's is $\sum_{i=0}^{K_r}\left(\begin{array}{c} N\\ i \end{array}\right)$ in the $r$-th repetition. 
Thus, it must hold that 
$
 \frac{M_r}{2}\leq \sum_{i=0}^{K_r}\left(\begin{array}{c} N\\ i \end{array} \right).
$
Since 
$
 \sum_{i=0}^{K_r}\left(\begin{array}{c} N\\ i \end{array} \right) \leq 2N^{K_r}
$, $K_r = \Omega\left(\frac{\log M_r}{\log N}\right)$ if $M_r \geq N$, otherwise $K_r \geq 1$. 
Therefore the number of the oracle calls is at most
$$
 \frac{9}{2}\sqrt{N}\log\log M \sum_{i=1}^{T'} \sqrt{\frac{\log N}{\log M_i}} + \frac{9}{2}\sqrt{N}\log\log M \log N,
$$
where the number of rows of $Z$ becomes $N$ or less after the $T'$-th repetition.
For $\{M_1,...,M_{T'}\}$, there exists a sequence of integers $\{k_1,...,k_{T'}\}$ $(1 \leq k_1 < \cdots < k_{T'}\leq \log M)$ such that 
$$
 1 \leq \frac{M}{2^{k_{T'}}} < M_{T'} \leq \frac{M}{2^{k_{T'-1}}} \leq \cdots \leq \frac{M}{2^{k_2}} < M_2 \leq \frac{M}{2^{k_1}} < M_1 = M
$$
since $M_r/2 \geq M_{r+1}$ for $r = 1,...,T'$.
Thus, we have
$$
 \sum_{i=1}^{T'} \frac{1}{\sqrt{\log M_i}} \leq \sum_{i=1}^{T'} \frac{1}{\sqrt{\log(M/2^{k_i})}}
\leq \sum_{i=0}^{\log{M}-1}\frac{1}{\sqrt{\log{M}-i}}
\leq 2\sqrt{\log M}.
$$
Then, the total number of the oracle calls is $O\left( \sqrt{N\log M \log N}\log\log M \right)$.

Next, we consider the success probability of our algorithm. By the analysis of the Grover search in \cite{BBHT96}, 
if the number of 1's of the current modified oracle is larger than $K_r$ in the $r$-th repetition, 
then we can find 1 in the current modified oracle with probability at least $1 - (3/4)^{\log\log M}$. 
This success probability worsens after $T$ rounds of repetition but still keeps a constant as follows: 
$
 ( 1 - (3/4)^{\log\log M} )^T \geq  ( 1 - 1/\log M)^{\log M} = \Omega(1).
$
\end{proof}

\begin{theorem}\label{N_times_M_L}
There is an OIP whose query complexity is $\Omega(\sqrt{\frac{N}{\log N}\log M})$. 
\end{theorem}
\begin{proof}
This can be shown in the same way as Theorem~5.1 in \cite{Amb02} as follows. 
Let $X$ be the set of all the oracles whose values are 1 at exactly $K$
positions and $Y$ be the set of all the oracles that have $1$'s at exactly $K+1$
positions. We consider the union of $X$ and $Y$ for 
our oracle identification problem. Thus, $M = |X| + |Y| = 
 \left(\begin{array}{c} N\\ K \end{array}\right) +
\left(\begin{array}{c} N\\ K+1 \end{array}\right)$, and therefore, we
have $\log M < K \log N$.
Let also a relation $R$ be the set of all $(f, f')$ such that $f \in
X$, $f' \in Y$ and they differ in exactly a single position. 
Then the parameters in Theorem~5.1 in \cite{Amb02} take values 
$m= \left(\begin{array}{c} N-K\\ 1 \end{array}\right) = N-K$, 
$m'= \left(\begin{array}{c} K+1\\ 1 \end{array}\right) = K+1$ and
$l = l' = 1$. Thus the lower bound is
$\Omega(\sqrt{(N-K)(K+1)})$. 
Since $\log M = O(K \log N)$, $K$ can be as large as $\Omega(\frac{\log
M}{\log N})$, which implies our lower bound.  
\end{proof}

Thus the bound in Theorem~1 is almost tight but not exactly. 
When $M = N$, however, we have another algorithm which is tight within a factor of constant. 
Although we prove the theorem for $M = N$, it also holds for $M = poly(N)$. 

\begin{theorem}\label{N_times_N}
The query complexity of any $N \times N$ OIP is $O(\sqrt{N})$.
\end{theorem}

\begin{proof}
Let $S$ be the given $N \times N$ matrix. Our algorithm is the following procedure: 

{\it Step 1.}~Let $Z = S$. 
If there is a column in $Z$ which has at least $\sqrt{N}$ $0$'s and at least $\sqrt{N}$ $1$'s, 
then perform a {\it classical} oracle call with this column. 
Eliminate all the inconsistent rows and update $Z$. 

{\it Step 2.}~Modify $Z$ to be $1$-sensitive. 
Perform the multi-target Grover search \cite{BBHT96} to obtain column $j$. 

{\it Step 3.}~Find a column $k$ which has $0$ and $1$ in some row while the column $j$ obtained in the Step 2 
has $1$ in that row (there must be such a column because any two rows are different). 
Perform a {\it classical} oracle call with column $k$ and remove inconsistent rows. 
Update $Z$. Repeat this step until $|Z| = 1$. 

Since the correctness of the algorithm is obvious, 
we only prove the complexity. 
A single iteration of Step 1 removes at least $\sqrt{N}$ rows, 
and hence we can perform at most $\sqrt{N}$ iterations 
(at most $\sqrt{N}$ oracle calls). 
Note that after this step each column of $Z$ has at most $\sqrt{N}$ $0$'s or at most $\sqrt{N}$ $1$'s. 
Since we perform the Column Flip in Step 2, we can assume that each column has at most $\sqrt{N}$ $1$'s. 
The Grover search in Step 2 needs $O(\sqrt{N})$ oracle calls. 
Since column $j$ has at most $\sqrt{N}$ $1$'s, 
the classical elimination in Step 3 needs at most $\sqrt{N}$ oracle calls.
\end{proof}

\section{Tight Upper Bounds for Small $M$}

In this section, we investigate the case that $M = N$ in more detail. 
Note that Theorem~3 is tight for the whole $N \times N$ OIP 
but not for its subfamilies. 
(For example, the Bernstein-Vazirani needs only $O(1)$ queries.) 
To seek optimal bounds for subfamilies, we introduce the following parameter: 
Let $S$ be an OIP given as an $N \times M$ matrix. 
Then $\#(S)$ be the maximum number of $1$'s in a single column of the matrix. 
We first give a lower bound theorem in terms of this parameter, 
which is a simplified version of Proposition~1. 

\begin{theorem}\label{lower_bound}
Let $S$ be an $N \times M$ matrix and $K = \#(S)$. 
Then $S$ needs
$\Omega(\sqrt{M/K})$ queries.
\end{theorem}

\begin{proof}  
Without loss of generality, we can assume that $S$ is $1$-sensitive, i.e., $K \le M/2$. 
We select $X$ ($Y$, resp.) as the upper (lower, resp.) half of $S$ (i.e., $|X| = |Y| = M/2$) 
and set $R = X \times Y$ (i.e., $(x,y) \in R$ for every $x \in X$ and $y \in Y$). 
Let $\delta_{j}$ be the number of 1's in the $j$-th column of $Y$. 
Now it is not hard to see that we can set $m=m'= \frac{M}{2}$, 
$l_{x,j}l_{y,j} = \max\{\delta_{j}(\frac{M}{2}-K_{j}+\delta_{j}), (\frac{M}{2}-\delta_{j})(K_{j}-\delta_{j})\}$ 
where $K_{j}$ is the number of $1$'s in column $j$. 
Since $K_{j} \le K$, this value is bounded from above by $\frac{M}{2}K$. 
Hence, Proposition~\ref{ambainis} implies 
$
\Omega\left(\sqrt{\frac{m m'}{l_{max}}}\right)\ge
\Omega\left(\sqrt{\frac{(\frac{M}{2})^2}{\frac{M}{2}K}}\right)=
\Omega\left(\sqrt{\frac{M}{K}}\right).
$
\end{proof}

Although this lower bound looks much simpler than Proposition~1, 
it is equally powerful for many cases. 
For example, we can obtain $\Omega(\sqrt{N})$ lower bound for the OIP given in Fig.~3 
which we denote by $X$. 
Note in general that if we need $t$ queries for a matrix $S$, 
then we also need at least $t$ queries for any $S^{'} \supseteq S$. 
Therefore it is enough to obtain a lower bound for the matrix $X^{'}$ which consists of the $N/2$ upper-half rows of $X$ 
and all the $1$'s of the right half can be changed to $0$'s by the Column Flip. 
Since $\#(X^{'}) = 1$, Theorem~4 gives us an lower bound of $\Omega(\sqrt{N})$. 

Now we give tight upper bounds for three subfamilies of $N \times N$ matrices. 
The first one is not a worst-case bound but an average-case bound: 
Let $AV(K)$ be an $N \times N$ matrix where each entry is $1$ with the probability $K/N$. 

\begin{theorem}\label{random}
The query complexity for $AV(K)$ is $\Theta(\sqrt{N/K})$ with 
high probability if $K = N^{\alpha}$ for $0 < \alpha <1$. 
\end{theorem}
\begin{proof}
Suppose that $X$ is an $AV(K)$. By using a standard Chernoff-bound argument,
we can show that the following three statements hold for $X$ with high probability
(Proofs are omitted).
(i) Let $c_i$ be the number of 1's in column $i$. Then for any $i$, $1/2 K \leq c_i \leq 2K$. 
(ii) Let $r_j$ be the number of 1's in row $j$. Then for any $j$, $1/2 K \leq r_j \leq 2K$.
(iii) Suppose that $D$ is a set of any $d$ columns in $X$ ($d$ is a function in $\alpha$ which is
constant since $\alpha$ is a constant). 
Then the number of rows which have 1's in all the columns in $D$ is at most $2\log N$.

Our lower bound is immediate from (i) by Theorem~\ref{lower_bound}. For the upper bound,
our algorithm is quite simple. Just perform the Grover search independently $d$ times. Each single
round needs $O(\sqrt{N/K})$ oracle calls by (ii). After that the number of candidates 
is decreased to $2\log N$ by (iii). Then we simply perform the classical elimination, 
just as step 3 of the algorithm in the proof of Theorem~3, 
which needs at most $2\log N$ oracle calls. Since $d$ is a constant, the overall complexity is
$O(\sqrt{N/K}) + \log N = \sqrt{N/K}$ if $K = N^\alpha$.
\end{proof}

The second subfamily is called a {\it balanced matrix}. 
Let $B(K)$ be a family of $N \times N$ matrices in which every row and every column has exactly $K$ $1$'s. 
(Again the theorem holds if the number of $1$'s is $\Theta(K)$.) 

\begin{theorem}\label{balanced}
The query complexity for $B(K)$ is $\Theta(\sqrt{N/K})$ if $K \le N^{1/3}$. 
\end{theorem}

\begin{proof}
The lower-bound part is obvious by Theorem~4. 
The upper-bound part is to use a single Grover search $+$ $K$ classical elimination. 
Thus the complexity is $O(\sqrt{N/K} + K)$, which is $O(\sqrt{N/K})$ if $K \le N^{1/3}$. 
\end{proof}

The third one is somewhat artificial. 
Let $H(k)$, called an {\it hybrid matrix} because it is a combination of Grover and Bernstein-Vazirani, 
be a matrix defined as follows: 
Let
$a=(a_1, a_2,\ldots, a_{n-k}, a_{n-k+1},\ldots, a_{n})$
and \\
$x=(x_1, x_2,\ldots, x_{n-k}, x_{n-k+1},\ldots, x_{n}).$ Then
$f_a(x)=1$ iff 
(i) $(a_1,..., a_{n-k})=(x_1,..., x_{n-k})$ and 
(ii) $(a_{n-k+1},...,a_{n})\cdot (x_{n-k+1},..., x_{n}) =0$ (mod $2$).
Fig.~4 shows the case that $k=2$ and $n=4$.

\begin{figure}[h]
\begin{center}
\resizebox{7.5cm}{!}{ 
\includegraphics*{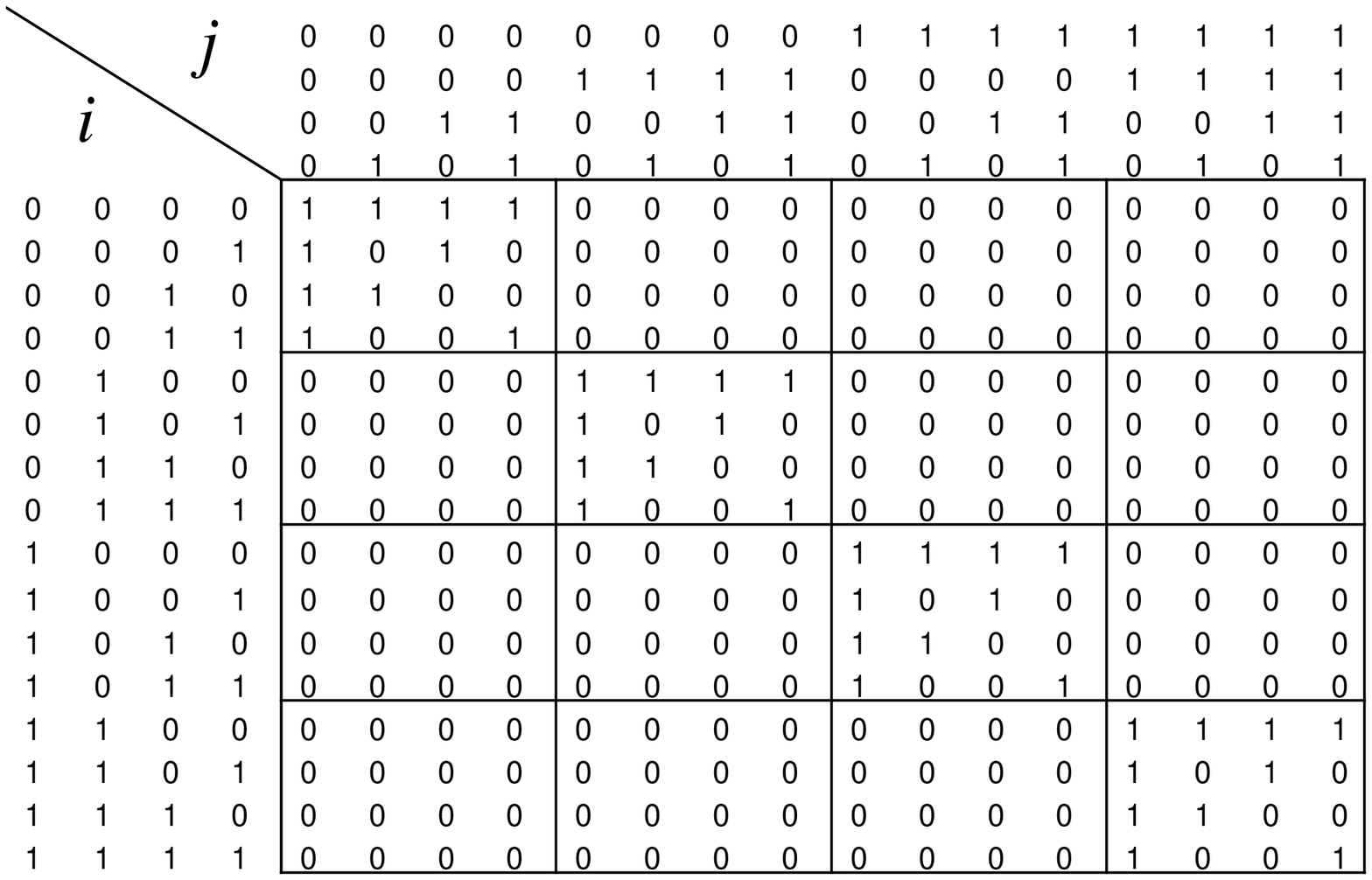} 
}
\caption{$H(k)$ with $n=4$ and $k=2$}
\end{center}
\end{figure}

\begin{theorem}\label{hybrid}
The query complexity for $H(k)$ is $\Theta(\sqrt{N/K})$, 
where $K = 2^{k}$.
\end{theorem}

\begin{proof}
We combine the Grover search \cite{Gro96,BBHT96} with BV
algorithm\cite{Bernstein97} to identify the oracle  $f_a$ 
by determining the hidden value $a$ of $f_a$. 
We first can determine the first $n-k$ bits of $a$. Fixing the
last $k$ bits to $|0 \rangle$, we apply the Grover search using oracle $f_a$
for the first $n-k$ bits to determine $a_1,...,a_{n-k}$. 
It should be noted that $f_{a}(a_1, \ldots, a_{n-k}, 0, \ldots, 0)
= 1$ and $f_{a}(x_1, \ldots, x_{n-k}, 0, \ldots, 0)=0$ for any
$x_1,...,x_k \neq a_1,...,a_k$. Next, we apply BV algorithm to determine the remaining $k$ bits of $a$. 
This algorithm requires $O(\sqrt{N/K})$ queries for the Grover search and 
$O(1)$ queries for BV algorithm to determine $a$.
Therefore we can identify the oracle $f_a$ using $O(\sqrt{N/K})$ 
queries.
\end{proof}

\section{Classical Lower and Upper Bounds}
The lower bound for the general $N \times M$ OIP is obviously $N$ if $M > N$. 
When $M = N$, we can obtain bounds being smaller than $N$ for some cases. 
\begin{theorem}
The deterministic query complexity for $N\times N$ OIP $S$ with $\#(S)=K$ is at least $\lfloor\frac{N}{K}\rfloor+\lfloor\log{K}\rfloor-2$.
\end{theorem}
\begin{proof}
Let $f_a$ be the current input oracle.
The following proof is due to the standard adversary argument. Let $A$ be any deterministic algorithm using the oracle $f_a$. 
Suppose that we determine $a \in \{0,1\}^n$ to identify the oracle $f_a$. Then the execution of $A$ is described as follows:
(i) In the first round, $A$ calls the oracle with the predetermined value $x_0$ and the oracle answers with $d_0=f_{a}(x_0)$. 
(ii) In the second round, $A$ calls the oracle with value $x_1$, which is determined by $d_0$ and the oracle answers with $d_1=f_{a}(x_1)$. 
(iii) In the $(i+1)$-st round, $A$ calls the oracle with $x_{i}$ which is determined by $d_0,d_1,...,d_{i-1}$ and 
the oracle answers with $d_{i} = f_{a}(x_{i})$.
(iv) In the $m$-th round $A$ outputs $a$ which is determined by
$d_0, d_1,..., d_{m-1}$ and stops. Thus, the execution of $A$ is
completely determined by the sequence $(d_0, d_1,..., d_{m-1})$
which is denoted by $A(a)$. (Obviously, if we fix a specific 
$a$, then $A(a)$ is uniquely determined).

Let $m_0 = \lfloor N/K \rfloor + \lfloor \log K \rfloor -3$ and suppose that $A$
halts in the $m_0$-th round. We compute the sequence
$(c_0,c_1,\ldots,c_{m_0}),\, c_i\in\{0,1\}$, and another sequence 
$(L_0,L_1,\ldots,L_{m_0}),\, L_i\subseteq \{a|a\in\{0,1\}^n\}$, as follows (note
that $c_0,\ldots,c_{m_0}$ are similar to $d_0,...,d_{m-1}$ above and are chosen by the adversary): (i) $L_0 = \{0,1\}^n$.
(ii) Suppose that we have already computed $L_0,...,L_i$, and $c_0,...,c_{i-1}$. Let $x_i$ be the value
with which $A$ calls the oracle in the $(i+1)$-st round.
(Recall that $x_i$ is determined by $c_0,...,c_{i-1}$.)
Let $L^0 = \{s\, |\, f_{s}(x_i) =0\}$ and $L^1 = \{s\, |\, f_{s}(x_i)=1\}$.
Then if $|L_i \cap L^0| \geq |L_i \cap L^1|$ then we set $c_i = 0$ and $L_{i+1} = L_i \cap L^0$.
Otherwise, i.e., if $|L_i\cap L^0|<|L_i\cap L^1|$, then we set $c_i = 1$ and $L_{i+1} = L_{i} \cap L^1$.

Now we can make the following two claims.

\textbf{Claim 1.} $|L_{m_0}|\geq 2$. 
(Reason: Note that $|L_0|=N$ and the size of $L_i$ decreases as $i$
increases. By the construction of $L_i$, one can see that until
$|L_i|$ becomes $2K$, its size decreases additively by at most $K$ in
a single round and after that it decreases multiplically at most one
half. The claim then follows by a simple calculation.) 

\textbf{Claim 2.} If $a \in L_{m_0}$, then $(c_0,\ldots ,c_{m_0})=A(a)$.
(Reason: Obvious since $a\in L_0\cap L_1\cap \cdots\cap L_{m_0}$.)

Now it follows that there are two different $a_1$ and $a_2$ in $L_{m_0}$ such that $A(a_1)=A(a_2)$
by Claims 1 and 2. Therefore $A$ outputs the same answer for two
different $a_1$ and $a_2$, a contradiction. 
\end{proof}

For the classical upper bounds, we only give the bound for the hybrid matrix. 
Similarly for $AV(K)$ and $B(K)$. 

\begin{theorem}
The deterministic query complexity for $H(k)$ is $O(\frac{N}{K}+\log{K})$.
\end{theorem}
\begin{proof} 
Let $f_a$ be the current input oracle. The algorithm consists of an exhaustive 
and a binary search to identify the oracle $f_a$ 
by determining the hidden value $a$ of $f_a$.
First, we determine the first $n-k$ bits of $a$ by fixing the last $k$ bits to all $0$'s and using exhaustive search. 
Second, we determine the last $k$ bits of $a$ by using binary search.
This algorithm needs $2^{n-k} (=\frac{N}{K})$ queries in the exhaustive search, 
and $O(k)(=O(\log{K}))$ queries in the binary search. 
Therefore, the total complexity of this algorithm is $O(2^{n-k}+ k) =O(\frac{N}{K}+\log{K})$. 
\end{proof}

\section{Concluding Remarks}
Some future directions are as follows: 
The most interesting one is a possible improvement of Theorem~1, 
for which our target is $O(\sqrt{N\log{M}})$. 
Also, we wish to have a matching lower bound, 
which is probably possible by making the argument of Theorem~2 a bit more exact. 
As mentioned before, in a certain situation, 
we do not have to determine the current oracle completely 
but have only to do that ``approximately", e.g., 
have to determine whether it belongs to some subset of oracles. 
It might be interesting to investigate how this approximation makes the problem easier 
(or basically not).

Most recently, it turned out that our problem OIP is equivalent to {\it exact learning},
which is a well-studied model of computional learning, by comments from Servedio.
\cite{GS01} has already shown interesting results on the quantum exact learning, which are 
independent of our main result on the quantum upper bound of any OIP. 
More precisely, \cite{GS01} defined a natural quantum version of two learning 
models and proved the equivalence up to polynomial factors between classical 
and quantum query complexity for the models.
Interpreting the result on the exact learning into the context of our OIP,
if there exists a quantum algorithm that solves an OIP $S$ with $Q$ queries
then there exists a deterministic algorithm that solves $S$ with $O(Q^3\log N)$ queries.

\subsection*{Acknowledgement}
The authors would like to thank Rocco Servedio for his comments on the relationships
between our problem and computational learning.

\end{document}